\documentstyle[12pt]{article}
\def\be{\begin{equation}}
\def\eeq{\end{equation}}
\def\bea{\begin{eqnarray}}
\def\eea{\end{eqnarray}}
\def\ora{\overrightarrow}
\def\ola{\overleftarrow}
\def\pd{\partial}
\def\ss{\footnotesize}
\def\id{{\rm l}\!{\rm l}}
\begin{document}
\title{Moyal Nahm Equations}
\author{Linda Baker$\footnote{e-mail: l.m.baker@durham.ac.uk}$\,  and David Fairlie$\footnote{e-mail: david.fairlie@durham.ac.uk}$\\
\\
Department of Mathematical Sciences,\\
         Science Laboratories,\\
         University of Durham,\\
         Durham, DH1 3LE, England}
\maketitle
\begin{abstract}
Various aspects of the Nahm equations in 3 and 7 dimensions are investigated.
The residues of the  variables at simple poles in the 7-dimensional case
form an algebra. A large class of matrix representations of this algebra is constructed. The large $N$ limit of these equations is taken by replacing the commutators by Moyal Brackets, and a set of non trivial solutions in a generalised  form of  Wigner distribution functions is obtained.  
\end{abstract}
\newpage
\section{Introduction}
When the study of Self Dual gauge fields was very fashionable, there was some interest in extending the theory to higher dimensions \cite{cor}\cite{gursey}\cite{ward}. In recent years interest in this subject has  been revived \cite{baulieu} partly because of the occurrence of Yang Mills gauge actions in the M(atrix) Theory approximation to string theories 
\cite{claudson}\cite{banks}\cite{vv}. The present article is a study of a class of solutions to the Nahm equations  \cite{craigie} in 7-dimensions, which are a particular form of the Self Dual Yang-Mills equations in Euclidean 8-dimensional space, where the gauge fields  $A^\mu,\ \ \mu=1\ldots 7$ depend only upon the 8th co-ordinate, $\tau$ and we work in a gauge where $A^8=0$. 
These equations take the form
\bea
\frac{\partial A^{1}}{\partial \tau}\,-\, [ A^{2},\ A^{7} ]\,-\, [ A^{6},\ A^{3} ]\,-\, [ A^{5},\ A^{4} ]\,=\,0\nonumber\\
\frac{\partial A^{2}}{\partial \tau}\,-\, [ A^{7},\ A^{1} ]\,-\, [ A^{5},\ A^{3} ]\,-\, [ A^{4},\ A^{6} ]\,=\,0\nonumber\\
\frac{\partial A^{3}}{\partial \tau}\,-\, [ A^{1},\ A^{6} ]\,-\, [ A^{2},\ A^{5} ]\,-\, [ A^{4},\ A^{7} ]\,=\,0\nonumber\\
\frac{\partial A^{4}}{\partial \tau}\,-\, [ A^{1},\ A^{5} ]\,-\, [ A^{6},\ A^{2} ]\,-\, [ A^{7},\ A^{3} ]\,=\,0\label{nahm}\\
\frac{\partial A^{5}}{\partial \tau}\,-\, [ A^{4},\ A^{1} ]\,-\, [ A^{3},\ A^{2} ]\,-\, [ A^{6},\ A^{7} ]\,=\,0\nonumber\\
\frac{\partial A^{6}}{\partial \tau}\,-\, [ A^{3},\ A^{1} ]\,-\, [ A^{2},\ A^{4} ]\,-\, [ A^{7},\ A^{5} ]\,=\,0{\nonumber}\\
\frac{\partial A^{7}}{\partial \tau}\,-\, [ A^{1},\ A^{2} ]\,-\, [ A^{3},\ A^{4} ]\,-\, [ A^{5},\ A^{6} ]\,=\,0\nonumber
\eea
The sum of the squares of the left hand sides of these equations  give the Lagrangian density for Yang Mills theory in 8-dimensions dependent only upon 
$\tau$ in the gauge $A^8=0$, up to divergence terms. This is just the Bogomol'nyi property of the equations (\ref{nahm}). Alternatively, the equations may be iterated to obtain the equations of motion for the Yang Mills field.
Note that when the $\tau$ dependence of $A^\mu$ is a simple pole, then the equations for the residues are just algebraic. The first objective of this paper is to initiate a study of the algebraic solution to such equations. It is known that a solution to (\ref{nahm}) takes the form
\begin{equation}
{A^i} \,=\,  \frac {-1}{2\tau}e_i\ \quad i=1\dots 7
\label{oct}
\end{equation}
where the $e_i$ form a basis for the imaginary octonions, but since these do not possess a matrix representation, as they are non-associative, one might wonder whether a matrix valued solution exists. Indeed many such solutions exist, and 
we determine the allowable $\tau$ dependence of some of them. We find a particular solution, modelled upon the  structure constants for octonionic multiplication and a more general solution in the form of a direct sum of $SU(2)$ representations.

The second objective of this paper is to consider the infinite limit of the     
Nahm equations in both 3 and 7-dimensions in the limit of large $N$ for the gauge group $SU(N)$. The motivation for this is a possible application to String Theory and Matrix Models. In this case the matrices $A^\mu$ go over to functions $X^\mu(x,p)$ on phase space   $(x,p)$. The matrix elements may be regarded as the Fourier components of $ X^\mu$. The commutator goes over to the Moyal Bracket
\cite{moyal}. This we recall is the antisymmetric part of the star product, which   acts on functions in phase space $(x,p)$. The star product of two functions $f(x,p)$ and $g(x,p)$ is defined as
\be
f(x,p) \star g(x,p) = f(x,p) e^{i \lambda (\ola{\partial_{x}} \ora{\partial_{p}}-\ola{\partial_{p}} \ora{\partial_{x}})} g(x,p)
\end{equation}
where $\lambda$ is a parameter.
The Moyal Bracket is proportional to the antisymmetric part of the star product and so the Moyal bracket of two functions $f(x,p)$ and $g(x,p)$ is written as
\be
\{ f,\ g \}_{MB} = \frac{1}{2 i} (f \star g\, - \,g \star f)
\end{equation}
It is the unique one parameter associative deformation of the Poisson Bracket
\cite{bayen}\cite{arveson}\cite{fletcher}. As $N=\frac{2\pi}{\lambda}$ passes through the odd integers the Moyal bracket $\{ X^{\mu},\ X^{\nu} \}_{MB}$
degenerates into an infinite direct sum of copies of the commutator 
$[A^\mu,\ A^\nu]$, where $A^\mu,\ A^\nu$ are $SU(N)$ matrices, the large $N$ limit of which is the Poisson Bracket
\be
\{ X^{\mu},\ X^{\nu} \}_{PB}\,=\, \frac{\pd X^\mu}{\pd x} \frac{\pd X^\mu}{\pd p}\,-\, \frac{\pd X^\mu}{\pd p} \frac{\pd X^\nu}{\pd x}.\label{poisson}
\end{equation} 
The square of this quantity is just the Schild \cite{schild}\cite{fai} form of the String Lagrangian, giving the same classical equations of motion as does the Nambu-Goto String. Thus the large $N$  limit of the Yang Mills Lagrangian in the strong coupling limit, where it is simply proportional to the square of the trace of the commutator, is equivalent to the String Lagrangian \cite{fai}. This may 
be viewed as a primitive form of a type of Maldacena conjecture \cite{maldacena}, relating a field theory in the large $N$ limit to a String Theory.  The phase space co-ordinates $x,\ p$ are to be interpreted as co-ordinate parametrisations on the world-sheet of the string. The idea had been advanced that the target space co-ordinates $X^\mu$ ($D_0$ branes ) may be represented by a generalised form of the Wigner function familiar from Quantum
Mechanics \cite{cur}\cite{fai2}. 
We shall demonstrate solutions of the Moyal-Nahm equations (where the commutators are replaced by Moyal brackets) which take the form of a generalised Wigner function.
\section{7-d Nahm equations}
The equations (\ref{nahm}) can be written more succinctly with the aid of totally antiysmmetric structure constants $C_{ijk}$ which, in fact, define the multiplication table of octonions of different index;
\be
e_i\times e_j\,=\,C_{ijk}e_k
\label{timestable}
\end{equation}
in the same was as the $\epsilon_{ijk}$ symbol does for the quaternions.
The Nahm equations  may then be written in the form
\be
\frac{\pd A_i}{\pd\tau}= \frac{1}{2}C_{ijk}[A_j,\ A_k]\label{Nahm2}
\end{equation}
with a solution for the residues $B_i$ of the simple poles in $\displaystyle{A_i=-\frac{1}{\tau}B_i}$ given by 
\be
B_i =C_{ijk}\label{mateixsol}
\end{equation}
An explicit set of matrices $B_i$ is listed in the appendix. No more general 
$\tau$ dependent solution of this type has been found. However, a very large class of solutions  whose matrices take the  form of a direct sum of representations
of the $SU(2)$ algebra, but which are nevertheless not reducible, with non-trivial $\tau$ dependence does exist and we proceed to explain their construction.
Let
\bea
B_1&=&-i\left(\begin{array}{ccc}\sigma_3&0&0\\ 0&\sigma_3&0\\ 0&0&\sigma_3\end{array}\right),\
B_2\,=\, -i\left(\begin{array}{ccc}\sigma_1&0&0\\ 0&b\sigma_2&0\\ 0&0&ic\sigma_3\end{array}\right),\ \nonumber\\
B_3&=&-i\left(\begin{array}{ccc}a\sigma_3&0&0\\ 0&\sigma_2&0\\ 0&0&ic\sigma_2\end{array}\right), \
B_4\,=\,-i\left(\begin{array}{ccc}ia\sigma_2&0&0\\ 0& b\sigma_3&0\\ 0&0&\sigma_2\end{array}\right),\ \nonumber\\
B_5&=&-i\left(\begin{array}{ccc}a\sigma_2&0&0\\ 0& -ib\sigma_3&0\\ 0&0&\sigma_1\end{array}\right),\ 
B_6\,=\,-i\left(\begin{array}{ccc}ia\sigma_3&0&0\\ 0& \sigma_1&0\\ 0&0&c\sigma_2\end{array}\right),\ \nonumber\\
B_7&=&-i\left(\begin{array}{ccc}\sigma_2&0&0\\ 0& ib\sigma_2&0\\ 0&0&c\sigma_3\end{array}\right).\label{diag}
\eea

It is somewhat surprising to find that this solution involves three arbitrary parameters, yet none of these parameters can be set to zero in such a way  that the solution remains faithful and irreducible.
Obviously many such solutions of direct sum form can be constructed where the $\sigma $s in each row can be replaced by representations of $SU(2)$ of arbitrary dimension. In particular, a representation of the Nahm algebra with $4\times4$ matrices can be found, by simply ommitting the last two rows in the above matrices (\ref {diag}). Our next task will be to search for the presence of solutions with non-trivial  $\tau$ dependence. 
An attempt was made to find a solution to the Nahm equations using the matrices $B_{i}$ (setting $a=1$, $b=1$, $c=1$). Each of these matrices was multiplied by a function of $\tau$, $f_{i}(\tau)$. Therefore,
\begin{equation}
A_{i}=f_{i}(\tau) B_{i}   \qquad i=1,\ldots,7
\end{equation}
However this ansatz is not sufficiently flexible and the previous result,
$f_i(\tau)\,=\,-\frac{1}{2 \tau}$ is all that is recovered.
A more general ansatz for a solution is to multiply each matrix $B_i$ by a diagonal matrix $C_i$ given by
\begin{equation}
C_i\,=\,\left(\begin{array}{ccc}f_i(\tau)\id&0&0\\ 0&g_i(\tau)\id&0\\ 0&0&h_i(\tau)\id\end{array}\right)\label{cdef}
\end{equation}
(This amounts to  multiplying each $\sigma$ matrix entry in each matrix $B_{i}$ by a different $\tau$ dependent function.)
It is easiest to consider each row of $\sigma$ matrices separately. First, we shall look at the top row of sigma matrices and put in the $\tau$ dependence by multiplying each matrix by a function $f_{i}(\tau)$.
Putting these 2 x 2 matrices into the Nahm equations gives the following set of differential equations
\bea
\frac{\pd f_{2}}{\pd \tau}&=&2 f_{1} f_{7}+2 f_{5} f_{3}-2 f_{4}f_{6}\nonumber\\
\frac{\pd f_{1}}{\pd \tau}&=&2 f_{2} f_{7},\ \quad \quad \frac{\pd f_{3}}{\pd \tau}=2 f_{2} f_{5},\ \quad \quad  \frac{\pd f_{4}}{\pd \tau}=2 f_{2} f_{6},\ \label{fred}\\
\frac{\pd f_{5}}{\pd \tau}&=&2 f_{2} f_{3},\ \quad \quad \frac{\pd f_{6}}{\pd \tau}=2 f_{2} f_{4},\ \quad \quad \frac{\pd f_{7}}{\pd \tau}=2 f_{2} f_{1}\nonumber
\eea
and the following constraints
\be
f_{7} f_{3}\,=\,f_{1} f_{5},\ \quad\quad \ f_{6} f_{7}\,=\,f_{1} f_{4},\ \quad\quad \ f_{3} f_{4}\,=\,f_{5} f_{6}.\label{fred2}
\end{equation}

Note that all of the differential equations involve $f_{2}$, but none of the constraints do.
These can be solved in terms of elliptic functions. It was found that
\bea
f_{6}\,=\,K_1 f_{3}\,=\, K_1 M_1 f_{1}=\frac{1}{2} K_1 M_1 Q_1 {\mathrm{ sn}} (q_1 \tau+d_1)\nonumber\\
f_{4}\,=\,K_1 f_{5}\,=\, K_1M_1 f_{7}=\frac{-i}{2} K_1 M_1 Q_1 {\mathrm{ cn}} (q_1 \tau+d_1)\label{fred3}\\
f_{2}\,=\,\frac{i}{2} q_1 {\mathrm{ dn}} (q_1 \tau+d_1)\nonumber
\eea
where cn, sn, dn are elliptic functions and $K_1$, $M_1$, $q_1$, $Q_1$, $d_1$ are all constants. The elliptic functions are related to each other by a parameter $k_1$ as follows
\bea
{\mathrm{ sn}}^{2}(x)+{\mathrm{ cn}}^{2}(x)=1\nonumber\\
{\mathrm{ dn}}^{2}(x)+k_1^{2} {\mathrm{ sn}}^{2}(x)=1\label{fred4}
\eea
where $k_1=\frac{Q_1}{q_1} \sqrt{1+M_1^{2}(1-K_1^{2})}$.\\
The following set of matrices solve Nahm's Equations

\bea
A_1&=&-i \left(\begin{array}{ccc}f_{1} \sigma_3&0&0\\ 0&g_{1} \sigma_3&0\\ 0&0&h_{1} \sigma_3\end{array}\right),\
A_2\,=\, -i \left(\begin{array}{ccc}f_{2}\sigma_1&0&0\\ 0&g_{2} \sigma_2&0\\ 0&0&i h_{2} \sigma_3\end{array}\right),\nonumber\\
A_3\,&=&\,-i \left(\begin{array}{ccc}f_{3}\sigma_3&0&0\\ 0&g_{3} \sigma_2&0\\ 0&0&i h_{3} \sigma_2\end{array}\right),\
A_4=-i \left(\begin{array}{ccc}i f_{4} \sigma_2&0&0\\ 0& g_{4} \sigma_3&0\\ 0&0&h_{4} \sigma_2\end{array}\right),\nonumber\\
A_5\,&=&\,-i \left(\begin{array}{ccc}f_{5}\sigma_2&0&0\\ 0& -i g_{5} \sigma_3&0\\ 0&0&h_{5} \sigma_1\end{array}\right) ,\
A_6=-i \left(\begin{array}{ccc}i f_{6} \sigma_3&0&0\\ 0& g_{6} \sigma_1&0\\ 0&0&h_{6} \sigma_2\end{array}\right),\nonumber\\
A_7\,&=&\,-i \left(\begin{array}{ccc}f_{7}\sigma_2&0&0\\ 0& i g_{7} \sigma_2&0\\ 0&0&h_{7} \sigma_3\end{array}\right)\label{fred5}
\eea

where the other $\tau$ dependent functions are given by
\bea
g_{5}\,=\,K_{2} g_{4}\,=\, K_{2} M_{2} g_{1}\,=\,\frac{1}{2} K_{2} M_{2} Q_{2} {\mathrm{ sn}} (q_{2} \tau+d_{2})\nonumber\\
g_{7}\,=\,K_{2} g_{2}\,=\, K_{2} M_{2} g_{3}\,=\,\frac{-i}{2} K_{2} M_{2} Q_{2} {\mathrm{ cn}} (q_{2} \tau+d_{2})\label{fred6}\\
g_{6}\,=\,\frac{i}{2} q_{2} {\mathrm{ dn}} (q_{2} \tau+d_{2})\nonumber
\eea
\bea
h_{6}\,=\,K_{3} h_{3}\,=\, K_{3} M_{3} h_{1}\,=\,\frac{1}{2} K_{3} M_{3} Q_{3} {\mathrm{ sn}} (q_{3} \tau+d_{3})\nonumber\\
h_{4}\,=\,K_{3} h_{5}\,=\, K_{3} M_{3} h_{7}\,=\,\frac{-i}{2} K_{3} M_{3} Q _{3} {\mathrm{ cn}} (q_{3} \tau+d_{3})\label{fred7}\\
h_{2}\,=\,\frac{i}{2} q_{3} {\mathrm{ dn}} (q_{3} \tau+d_{3}).\nonumber
\eea

\section{Moyal-Nahm Equations}
Consider a field $X^{k}$ ($k$=0,1,2,3) in four dimensions where $X^{k}$ depends  upon only one co-ordinate (in this case $t$) and phase space $(x,p)$. The gauge is fixed so $X^{0}$ is a constant.
The Moyal Nahm equations in three dimensions are:
\bea
\frac{\partial X^{1}}{\partial t}=\{ X^{2},X^{3} \}_{MB} \nonumber\\
\frac{\partial X^{2}}{\partial t}=\{ X^{3},X^{1} \}_{MB} \\
\frac{\partial X^{3}}{\partial t}=\{ X^{1},X^{2} \}_{MB} \nonumber
\eea
If the Moyal brackets were to be replaced by commutators and the functions $X^{k}(t,x,p)$ were  replaced by matrices $X^{k}(t)$ then the equations would become the Nahm equations for a self dual field. 
The main idea  to solve these\hfill\break Moyal-Nahm equations is to use the following ansatz:
\be
X^{i}=i \int_{-\infty}^{\infty} \psi_{j} ^{\dag} (x-y,t)\epsilon ^{ijk} \psi_{k} (x+y,t) e^
{2 \pi i p y/\lambda}  {\mathrm{d}} y
\end{equation} 
where  $\psi (x,t)$  are three component wave functions. These wave functions were chosen to be of the following form:
\begin{equation}
\psi(x,t)\,=\,\left( \begin{array}{c}
	     \psi_{1}(x,t) \\
	     \psi_{2}(x,t) \\
	     \psi_{3}(x,t) 
	\end{array} \right)
  \,=\,\left( \begin{array}{c}
	     f_{1}(t) \phi_{1}(x) \\
	     f_{2}(t) \phi_{2}(x) \\
	     f_{3}(t) \phi_{3}(x) 
	\end{array} \right)\label{annsatz}
\end{equation}		
where the $\phi_i(x)$ are orthonormal wave functions.
The star product of $X^{j}$ and $X^{k}$ is calculated as follows:
{\setlength\arraycolsep{2pt}
\begin{eqnarray}
X^{j} \star X^{k}=-\int\!\!\!\int &&\psi_{i}^{\dag}(x-y,t) \epsilon{} ^{jil} \psi_{l} (x+y,t) e^
{2 \pi i p y/\lambda} \star {}  \nonumber\\ && {} \psi_{m}^{\dag} (x-y',t)\epsilon ^{kmn} \psi_{n} (x+y',t) e^
{2 \pi i p y'/\lambda} {\mathrm{d}} y {\mathrm{d}} y'\nonumber 
\eea
\bea
=-\int\!\!\!\int &&\psi_{i}^{\dag}(x-y+y',t) \epsilon ^{jil}\psi_{l} (x+y+y',t) e^
{2 \pi i p y/\lambda} \nonumber\\ && \psi_{m}^{\dag} (x-y'-y,t)\epsilon ^{kmn} \psi_{n} (x+y'-y,t) e^
{2 \pi i p y'/\lambda} {\mathrm{d}} y {\mathrm{d}} y' \nonumber
\eea
\bea
=-\frac{1}{2} \int \psi_{m} ^{\dag} (x-y,t)\epsilon ^{kmn} Z(t) \epsilon ^{jil} \psi_{l} (x+y,t) e^{2 \pi i p y/\lambda}  {\mathrm{d}} y
\eea
\bea
=-\frac{1}{2} \int \phi_{m} ^{\dag} (x-y) f^{\dag}_{ms}(t) \epsilon ^{ksn} Z(t) \epsilon ^{jir} f_{rl}(t) \phi_{l} (x+y) e^{2 \pi i p y/\lambda}  {\mathrm{d}} y\nonumber
\eea
where orthogonality of the  $\phi_{k}(x)$ is assumed to be of the form
\be
\int_{-\infty}^{\infty} \phi_{j}^{\dag}(x) \phi_{k}(x) {\mathrm{d}} x = \delta_{jk}
\end{equation}
and 
\be
Z(t)=f^{\dag} f \qquad \textrm{where } f=\left( \begin{array}{ccc}
	    					 f_{1}(t) & 0 & 0 \\
	   					 0 & f_{2}(t) & 0 \\
	 					 0 & 0 & f_{3}(t) 
					\end{array} \right).
\end{equation} 
The partial derivative $\frac{\partial X^{i}}{\partial t}$ can be written as
\be
\frac{\partial X^{i}}{\partial t}=i \int \phi_{j}  ^{\dag} (x-y) \frac{\partial }{\partial t}(f^{\dag}(t) \epsilon ^{ijk} f(t)) \phi_{k} (x+y) e^
{2 \pi i p y/\lambda}  {\mathrm{d}} y.
\end{equation}

By putting these into the Moyal-Nahm equations one obtains three matrix equations of the form
\be
i \frac{\partial }{\partial t}(f^{\dag}(t) \epsilon ^{1} f(t))=\frac{-1}{4 i} ( f^{\dag}(t) \epsilon ^{3} Z(t) \epsilon ^{2}  f(t)-f^{\dag}(t) \epsilon ^{2} Z(t) \epsilon ^{3}  f(t))
\end{equation} 
where $\epsilon ^{i}$ is a 3 x 3 matrix with $jk^{\mathrm{th}}$ entry $\epsilon ^{ijk}$.
Equating the entries in the matrices gives differential equations of the form
\bea
\frac{\partial}{\partial t} (f_{2}^{*} f_{3}^{ })\,=\,-\frac{1}{4} |f_{1}|^{2} (f^{*}_{2} f_{3}^{ })\nonumber\\
\frac{\partial}{\partial t} (f^{*}_{3} f_{2}^{ })\,=\,-\frac{1}{4} |f_{1}|^{2} (f^{*}_{3} f_{2}^{ })
\eea
and cycle combinations of these. 
These can be used to create the following set of three differential equations.
\bea
\frac{\partial}{\partial t} (|f_{2}|^{2} |f_{3}|^{2})\,=\,-\frac{1}{2} |f_{1}|^{2} |f_{2}|^{2} |f_{3}|^{2} \nonumber\\ 
\frac{\partial}{\partial t} (|f_{3}|^{2} |f_{1}|^{2})\,=\,-\frac{1}{2} |f_{1}|^{2} |f_{2}|^{2} |f_{3}|^{2} \\ 
\frac{\partial}{\partial t} (|f_{1}|^{2} |f_{2}|^{2})\,=\,-\frac{1}{2} |f_{1}|^{2} |f_{2}|^{2} |f_{3}|^{2}. \nonumber
\eea 
Note that for each of the above, the right hand side of the equations is always the same.

\subsection {Simplest Solution}
The simplest solution is to set all the $f_{i}$ equal to each other. This gives the solution
\be
|f_{1}|^{2}\,=\,|f_{2}|^{2}\,=\,|f_{3}|^{2}\,=\,\frac{4}{t+K}
\end{equation}
so that
\be
f_{1}(t)\,=\,f_{2}(t)\,=\,f_{3}(t)\,=\,\frac{2}{\sqrt{t+K}}
\end{equation}
where $K$ is an arbitrary constant.

\subsection {Another Simple Solution}
By setting two of the $f_{i}$ equal to each other then a solution in terms of the hyperbolic functions can be found.
\bea
|f_{1}|^{2}\,=\,|f_{2}|^{2}&=&4 q \coth (q t+K)\nonumber\\
|f_{3}|^{2}&=&8 q \,{\mathrm{ csch}} (2 q t+2 K) 
\eea
so that
\bea
f_{1}(t)\,=\,f_{2}(t)&=&2 \sqrt{q \coth (q t+K)}\nonumber\\
f_{3}(t)&=&2 \sqrt{2 q \,{\mathrm{ csch}} (2 q t+2 K)}
\eea
where $K$ and $q$ are both real constants.

\subsection {General Solution}
However, ideally we want a general solution to these equations. In this case the solutions are written in terms of elliptic functions sn, cn and dn. The most general solution was found to be:

\bea
& &\lefteqn{|f_{1}|^{2}\,=\,4 q k  {\mathrm{ sn}} (q t+c)}\nonumber\\
& &|f_{2}|^{2}\,=\, 2 q k {\mathrm{ sn}} (q t+c) + \frac{2 q ({\mathrm{ dn}} (q t+c) {\mathrm{ cn}} (q t+c)+1)}{{\mathrm{ sn}} (q t+c)}\\
& &|f_{3}|^{2}\,=\,2 q k {\mathrm{ sn}} (q t+c) + \frac{2 q ({\mathrm{ dn}} (q t+c) {\mathrm{ cn}} (q t+c)-1)}{{\mathrm{ sn}} (q t+c)}\nonumber
\eea
$k$, $q$, and $c$ are all constants but may have to be carefully chosen in order to ensure that all the $|f_{i}|^{2}$ are positive. k depends on the elliptic functions. \\
\\A more aesthetically pleasing  form of solution is as follows:
\bea
& &|f_{1}|^{2}\,=\,4 q k^{2} \frac{{\mathrm{ sn}} (q t+c) {\mathrm{ cn}} (q t+c)}{{\mathrm{ dn}} (q t+c)}\nonumber\\
& &|f_{2}|^{2}\,=\,-4 q \frac{{\mathrm{ cn}} (q t+c) {\mathrm{ dn}} (q t+c)}{{\mathrm{ sn}} (q t+c)}\label{eps}\\
& &|f_{3}|^{2}\,=\,4 q \frac{{\mathrm{ dn}} (q t+c) {\mathrm{ sn}} (q t+c)}{{\mathrm{ cn}} (q t+c)}\nonumber
\eea
$q$,$c$ and $k$ are all constants. Again, $k$ depends on the elliptic functions.

\section{7D Moyal Equations}
The Moyal Nahm equations in seven dimensions are:
\bea
\frac{\partial X^{1}}{\partial t}=\{ X^{2},X^{7} \}_{MB}+\{ X^{6},X^{3} \}_{MB}+\{ X^{5},X^{4} \}_{MB}\nonumber\\
\frac{\partial X^{2}}{\partial t}=\{ X^{7},X^{1} \}_{MB}+\{ X^{5},X^{3} \}_{MB}+\{ X^{4},X^{6} \}_{MB}\nonumber\\
\frac{\partial X^{3}}{\partial t}=\{ X^{1},X^{6} \}_{MB}+\{ X^{2},X^{5} \}_{MB}+\{ X^{4},X^{7} \}_{MB}\nonumber\\
\frac{\partial X^{4}}{\partial t}=\{ X^{1},X^{5} \}_{MB}+\{ X^{6},X^{2} \}_{MB}+\{ X^{7},X^{3} \}_{MB}\label{moynam}\\
\frac{\partial X^{5}}{\partial t}=\{ X^{4},X^{1} \}_{MB}+\{ X^{3},X^{2} \}_{MB}+\{ X^{6},X^{7} \}_{MB}\nonumber\\
\frac{\partial X^{6}}{\partial t}=\{ X^{3},X^{1} \}_{MB}+\{ X^{2},X^{4} \}_{MB}+\{ X^{7},X^{5} \}_{MB\nonumber}\\
\frac{\partial X^{7}}{\partial t}=\{ X^{1},X^{2} \}_{MB}+\{ X^{3},X^{4} \}_{MB}+\{ X^{5},X^{6} \}_{MB}\nonumber
\eea
The set of matrices $B_{i}$ can also be used to find a solution to the equations
(\ref{moynam}).Using the ansatz
\be
A_{i}=i \int_{-\infty}^{\infty} \psi ^{\dag} (x-y,\tau) B_{i} \psi (x+y,\tau) e^
{2 \pi i p y/\lambda}  {\mathrm{d}} y\label{anz1}
\end{equation} 
where $\psi (x,t)$  are six component wave functions of the form
\be
\psi_j= f_j(\tau)\phi(x)_j,\ \ j\ {\rm not\ summed}\label{ann2}
\end{equation}
we find a rather simple solution in terms of this ansatz of the form
\bea
&&\!\!\!\{f_1,f_2,f_3,f_4,f_5,f_6\}\,=\nonumber\\
&&
\!\!\!\{ \frac{2 \sqrt{K_{1}} e^{i \theta_{1}}}{\sqrt{1-e^{-K_{1} \tau}}},\frac{2 \sqrt{K_{1}} e^{i \theta_{2}}}{\sqrt{e^{K_{1} \tau}-1}},\frac{2 \sqrt{K_{2}} e^{i \theta_{3}}}{\sqrt{1-e^{-K_{2} \tau}}}, \frac{2 \sqrt{K_{2}} e^{i \theta_{4}}}{\sqrt{e^{K_{2} \tau}-1}}, \frac{2 \sqrt{K_{3}} e^{i \theta_{5}}}{\sqrt{1-e^{-K_{3} \tau}}},\frac{2 \sqrt{K_{3}} e^{i \theta_{6}}}{\sqrt{e^{K_{3} \tau}-1}}\}
\nonumber
\eea
All $K_{i}$ and $\theta_{j}$ are constants.

The previous solution used the two-dimensional $\sigma$ matrix representation of $SU(2)$. If instead the three dimensional representation of $SU(2)$, which involves the completely antisymmetric matrices $\epsilon_{ijk}$, is used to construct the $B_i$ a more general $\tau$ dependent solution can be found along similar lines to that in section 3. This time the matrices $B_i$ were taken to be
\bea
B_1&=&-\left(\begin{array}{cc}\epsilon_3&0\\ 0&\epsilon_3\end{array}\right),\
B_2= -\left(\begin{array}{cc}\epsilon_1&0\\ 0&\epsilon_2\end{array}\right),\ 
B_3=-\left(\begin{array}{cc}\epsilon_3&0\\ 0&\epsilon_2\end{array}\right), \
B_4=-\left(\begin{array}{cc}i\epsilon_2&0\\ 0& \epsilon_3\end{array}\right),\ \nonumber\\
B_5&=&-\left(\begin{array}{cc}\epsilon_2&0\\ 0& i\epsilon_3\end{array}\right),\ 
B_6=-\left(\begin{array}{cc}i\epsilon_3&0\\ 0& \epsilon_1\end{array}\right),\ B_7=-\left(\begin{array}{cc}\epsilon_2&0\\ 0& i\epsilon_2\end{array}\right).\label{diag2}
\eea
where the $jk^{\mathrm{th}}$ entry of the matrix $\epsilon_{i}$ is given by the totally antisymmtric tensor  $\epsilon_{ijk}$. 

The same ansatz (\ref{anz1})(\ref{ann2}) was used as before but with the new $B_i$ matrices. The $\tau$ dependent functions $f_{i}$ were found to be of the same form as the solution to the 3d Moyal Nahm equations when solved using $\epsilon$ matrices (\ref{eps}). They are as follows

\bea
& &|f_{1}|^{2}=4 q k^{2} \frac{{\mathrm{ sn}} (q \tau+c) {\mathrm{ cn}} (q \tau+c)}{{\mathrm{ dn}} (q \tau+c)},\,\,|f_{4}|^{2}=4 Q K^{2} \frac{{\mathrm{ sn}} (Q \tau+b) {\mathrm{ cn}} (Q \tau+b)}{{\mathrm{ dn}} (Q \tau+b)}\nonumber\\
& &|f_{2}|^{2}=-4 q \frac{{\mathrm{ cn}} (q \tau+c) {\mathrm{ dn}} (q \tau+c)}{{\mathrm{ sn}} (q \tau+c)},\,\, |f_{5}|^{2}=-4 Q \frac{{\mathrm{ cn}} (Q \tau+b) {\mathrm{ dn}} (Q \tau+b)}{{\mathrm{ sn}} (Q \tau+b)}\nonumber\\
& &|f_{3}|^{2}=4 q \frac{{\mathrm{ dn}} (q \tau+c) {\mathrm{ sn}} (q \tau+c)}{{\mathrm{ cn}} (q \tau+c)}, \,\quad |f_{6}|^{2}=4 Q \frac{{\mathrm{ dn}} (Q \tau+b) {\mathrm{ sn}} (Q \tau+b)}{{\mathrm{ cn}} (Q \tau+b)}.
\eea
These solutions can be extended for direct sums of more than two $\epsilon_{i}$ matrices.

\section*{Acknowledgement}
Linda Baker is grateful to EPSRC for a postgraduate research award. David Fairlie wishes to thank Tatsuya Ueno for discussions.

\newpage
\section*{Appendix}
The following 7 x 7 matrices which solve the Nahm algebra are created using the octonionic structure constants $c_{ijk}$ which are taken to be:

\begin{equation}
c_{127}=c_{631}=c_{541}=c_{532}=c_{246}=c_{347}=c_{567}=1
\end{equation}
These are totally antisymmetric. All other $c_{ijk}$ are zero.
The $jk^{th}$ entry of the matrix $m_{i}$ is given by $[m_{i}]_{jk}=c_{ijk}$.

\begin{eqnarray}
m_{1}={\ss
\left(\begin{array}{ccccccc}
0 & 0 & 0 & 0 & 0 & 0 & 0\\
0 & 0 & 0 & 0 & 0 & 0 & 1\\
0 & 0 & 0 & 0 & 0 & -1 & 0\\
0 & 0 & 0 & 0 & -1 & 0 & 0\\
0 & 0 & 0 & 1 & 0 & 0 & 0\\
0 & 0 & 1 & 0 & 0 & 0 & 0\\
0 & -1 & 0 & 0 & 0 & 0 & 0
\end{array}\right)}&,& \  
m_{2}=
{\ss\left(\begin{array}{ccccccc}
0 & 0 & 0 & 0 & 0 & 0 & -1\\
0 & 0 & 0 & 0 & 0 & 0 & 0\\
0 & 0 & 0 & 0 & -1 & 0 & 0\\
0 & 0 & 0 & 0 & 0 & 1 & 0\\
0 & 0 & 1 & 0 & 0 & 0 & 0\\
0 & 0 & 0 & -1 & 0 & 0 & 0\\
1 & 0 & 0 & 0 & 0 & 0 & 0 
\end{array}\right)},\  \nonumber\\ 
m_{3}={\ss
\left(\begin{array}{ccccccc}
0 & 0 & 0 & 0 & 0 & 1 & 0\\
0 & 0 & 0 & 0 & 1 & 0 & 0\\
0 & 0 & 0 & 0 & 0 & 0 & 0\\
0 & 0 & 0 & 0 & 0 & 0 & 1\\
0 & -1 & 0 & 0 & 0 & 0 & 0\\
-1 & 0 & 0 & 0 & 0 & 0 & 0\\
0 & 0 & 0 & -1 & 0 & 0 & 0
\end{array}\right)}&,& \
m_{4}={\ss
\left(\begin{array}{ccccccc}
0 & 0 & 0 & 0 & 1 & 0 & 0\\
0 & 0 & 0 & 0 & 0 & -1 & 0\\
0 & 0 & 0 & 0 & 0 & 0 & -1\\
0 & 0 & 0 & 0 & 0 & 0 & 0\\
-1 & 0 & 0 & 0 & 0 & 0 & 0\\
0 & 1 & 0 & 0 & 0 & 0 & 0\\
0 & 0 & 1 & 0 & 0 & 0 & 0
\end{array}\right)}, \  \nonumber\\ 
m_{5}={\ss
\left(\begin{array}{ccccccc}
0 & 0 & 0 & -1 & 0 & 0 & 0\\
0 & 0 & -1 & 0 & 0 & 0 & 0\\
0 & 1 & 0 & 0 & 0 & 0 & 0\\
1 & 0 & 0 & 0 & 0 & 0 & 0\\
0 & 0 & 0 & 0 & 0 & 0 & 0\\
0 & 0 & 0 & 0 & 0 & 0 & 1\\
0 & 0 & 0 & 0 & 0 & -1 & 0
\end{array}\right)}&,& \ 
m_{6}={\ss
\left(\begin{array}{ccccccc}
0 & 0 & -1 & 0 & 0 & 0 & 0\\
0 & 0 & 0 & 1 & 0 & 0 & 0\\
1 & 0 & 0 & 0 & 0 & 0 & 0\\
0 & -1 & 0 & 0 & 0 & 0 & 0\\
0 & 0 & 0 & 0 & 0 & 0 & -1\\
0 & 0 & 0 & 0 & 0 & 0 & 0\\
0 & 0 & 0 & 0 & 1 & 0 & 0
\end{array}\right)},\nonumber\\
\quad\quad m_{7}={\ss
\left(\begin{array}{ccccccc}
0 & 1 & 0 & 0 & 0 & 0 & 0\\
-1 & 0 & 0 & 0 & 0 & 0 & 0\\
0 & 0 & 0 & 1 & 0 & 0 & 0\\
0 & 0 & -1 & 0 & 0 & 0 & 0\\
0 & 0 & 0 & 0 & 0 & 1 & 0\\
0 & 0 & 0 & 0 & -1 & 0 & 0\\
0 & 0 & 0 & 0 & 0 & 0 & 0
\end{array}\right)}&.&\nonumber
\end{eqnarray}

\end{document}